\documentclass[conference]{IEEEtran}
\IEEEoverridecommandlockouts
\usepackage{cite}
\usepackage{amsmath,amssymb,amsfonts}
\usepackage{algorithmic}
\usepackage{graphicx}
\usepackage{comment}
\usepackage{textcomp}
\usepackage{listings}
\usepackage{xcolor}
\usepackage{booktabs}
\usepackage{pgfplots}
\pgfplotsset{compat=newest}
\usepackage{pgfplotstable}
\def\BibTeX{{\rm B\kern-.05em{\sc i\kern-.025em b}\kern-.08em
    T\kern-.1667em\lower.7ex\hbox{E}\kern-.125emX}}

\author{
Tushar Vatsa\textsuperscript{*}, Vibha Belavadi\textsuperscript{*}, Priya Shanmugasundaram\textsuperscript{*}, Suhas Suresha\textsuperscript{*}, Dewang Sultania \\
Adobe Inc. \\
345 Park Avenue, San Jose, CA 95110
\thanks{*These authors contributed equally to this work.}
}

\begin{document}

\title{FUSE: Failure-aware Usage of Subagent Evidence for MultiModal Search and Recommendation}


\maketitle

\begin{abstract}
Multimodal creative assistants decompose user goals and route tasks to subagents for layout, styling, retrieval, and generation. Retrieval quality is pivotal, yet failures can arise at several stages: understanding user intent, choosing content types, finding candidates (recall), or ranking results. Meanwhile, sending and processing images is costly, making naive multimodal approaches impractical. We present FUSE: Failure-aware Usage of Subagent Evidence for multimodal search and recommendation. FUSE replaces most raw-image prompting with a compact Grounded Design Representation (GDR): a selection-aware JSON of canvas elements (image, text, shape, icon, video, logo), structure, styles, salient colors, and user selection provided by the Planner team. FUSE implements seven context budgeting strategies: comprehensive baseline prompting, context compression, chain-of-thought reasoning, mini-shot optimization, retrieval-augmented context, two-stage processing, and zero-shot minimalism. Finally, a pipeline attribution layer monitors system performance by converting subagent signals into simple checks: intent alignment, content-type/routing sanity, recall health (e.g., zero-hit and top-match strength), and ranking displacement analysis. We evaluate the seven context budgeting variants across 788 evaluation queries from diverse users and design templates (refer Figure~\ref{fig:contextual_search}). Our systematic evaluation reveals that \textbf{Context Compression achieves optimal performance across all pipeline stages}, with 93.3\% intent accuracy, 86.8\% routing success(with fallbacks), 99.4\% recall, and 88.5\% NDCG@5. This approach demonstrates that strategic context summarization outperforms both comprehensive and minimal contextualization strategies. The narrow intent performance variance (89.5-93.3\%) across variants validates that all context budgeting approaches provide substantial benefits over zero-shot baselines. We also measure p50/p95/p99 latency and normalized token cost: Context Compression delivers the best latency–cost trade-off (45\% lower p95 vs. Baseline; 8 times fewer input tokens vs. Chain-of-Thought) while Chain-of-Thought provides maximal reasoning at the highest cost.
\end{abstract}

\begin{IEEEkeywords}
creative assistants, pipeline attribution, subagent evidence, graphical design representation, context budgeting
\end{IEEEkeywords}

\begin{figure*}[t]
\centering
\includegraphics[width=\textwidth]{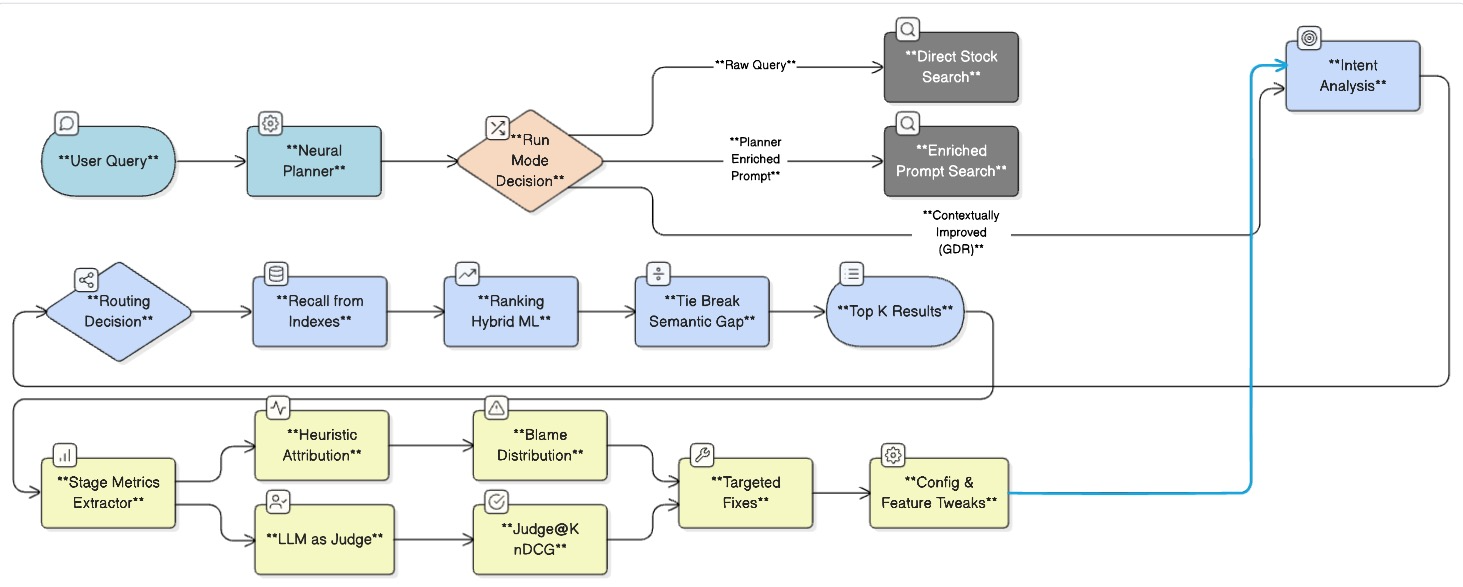}
\caption{Multi‑stage search system flow with FUSE integration. The diagram shows three operation modes: (a) raw query, (b) superagent‑enriched prompt, and (c) superagent prompt + compact GDR. The contextual subagent performs intent analysis, content‑type routing, recall, ranking, and semantic tie‑breaking. The evaluation and attribution layer extracts stage‑wise metrics, applies heuristic and LLM‑as‑judge evaluation, localizes failures via blame distribution, and drives targeted fixes and configuration updates.}
\label{fig:architecture}
\end{figure*}

\section{Introduction}
Traditional multimodal search and recommendation in creative platforms is largely non‑assistive: users condense intent into broad keywords to fetch assets, skim through presented results to select the asset appropriate for their intent, decide the modality or tool to be used, all while understanding different platform specific features. This burdens both novices and experts often leading to missed retrievals despite the platform possessing assets that might have been able to satisfy the user intent if the user experiences were more assistive. The advent of large language models has enabled customers to converse in natural language with various systems on the web, increasing their expectations from traditional search and recommendation systems to scale to natural language inputs and improve ease of use through real-time discovery assistants. 

Modern multimodal assistants address this by adopting agentic orchestration: a neural planner decomposes user queries and routes tasks to specialized subagents for layout, styling, retrieval, and generation~\cite{yao2023react,schick2023toolformer,liu2023cascades}. Within this architecture, the search subagent is a linchpin; downstream actions cannot redeem poor retrieval. Failures can originate anywhere in the pipeline (intent, routing, recall, ranking), and naïvely sending rich multimodal context—especially images not only degrades the output quality but also significantly increases latency and token cost~\cite{shi2024agent,zhang2024dynamic}. 

Vision-language models present significant overhead in terms of latency and cost, as each image must be tiled into patches (e.g., $512\times512$) and contributes hundreds of tokens in addition to a fixed per-image overhead~\cite{spurnow2024pricing}. Commercial APIs price vision input at roughly \$10 per million input tokens and \$40 per million output tokens, with high-quality image generation reaching \$0.17 per image~\cite{openai2024pricing}. Latency is also substantial: benchmarks of GPT-4o (Vision) report over $5$ seconds per $800\times800$ image, compared to hundreds of milliseconds for task-specific detectors~\cite{aimultiple2024vlm}. Optimized vision-language models like FastVLM reduce but do not eliminate this gap, achieving up to $85\times$ faster time-to-first-token than LLaVA-OneVision while retaining accuracy~\cite{jiang2024fastvlm}. These drawbacks make naive raw-image prompting impractical for production grade pipelines that must meet strict real-time SLAs and cost budgets.

\textbf{FUSE: Failure‑aware Usage of Subagent Evidence.} We propose FUSE, a production grade recipe for multimodal search and recommendation that unifies (i) a compact \emph{Graphical Design Representation} (GDR) from planner for conditioning, (ii) \emph{budgeted context policies} that right‑size prompts under cost caps, and (iii) a \emph{pipeline attribution} layer that detects failure modes and selects/escalates policies accordingly.

We experimented with three practical modes that reflect how real users engage the system (shown in Figure \ref{fig:architecture}):
\begin{enumerate}
  \item \emph{Raw query} $\rightarrow$ direct search with no additional context.
  \item \emph{Planner‑enriched prompt} $\rightarrow$ the neural planner expands and clarifies the user request.
  \item \emph{GDR + planner-enriched prompt} $\rightarrow$ the planner's prompt is augmented with GDR.
\end{enumerate}

 Over these modes, FUSE applies lightweight prompting variants chosen to meet latency and token constraints: adjustable examples (none, brief, or retrieved), GDR compression for large contexts, selective reasoning triggered only by ambiguity, and two-stage processing. Policies are chosen on demand rather than using fixed configurations.

A stage‑wise attribution layer converts subagent evidence into operational checks: (1) \emph{intent alignment} (semantic proximity between the planner subprompt and expected intent), (2) \emph{routing sanity} (content‑type confidence with fallbacks), (3) \emph{recall health} (zero‑hit and top‑match strength), and (4) \emph{ranking stability}. These checks serve two purposes: they enable real-time policy adjustments within budget constraints and inform offline system tuning (such as fallback mappings and re-ranking thresholds).

\begin{figure*}[t]
\centering
\includegraphics[width=0.48\textwidth]{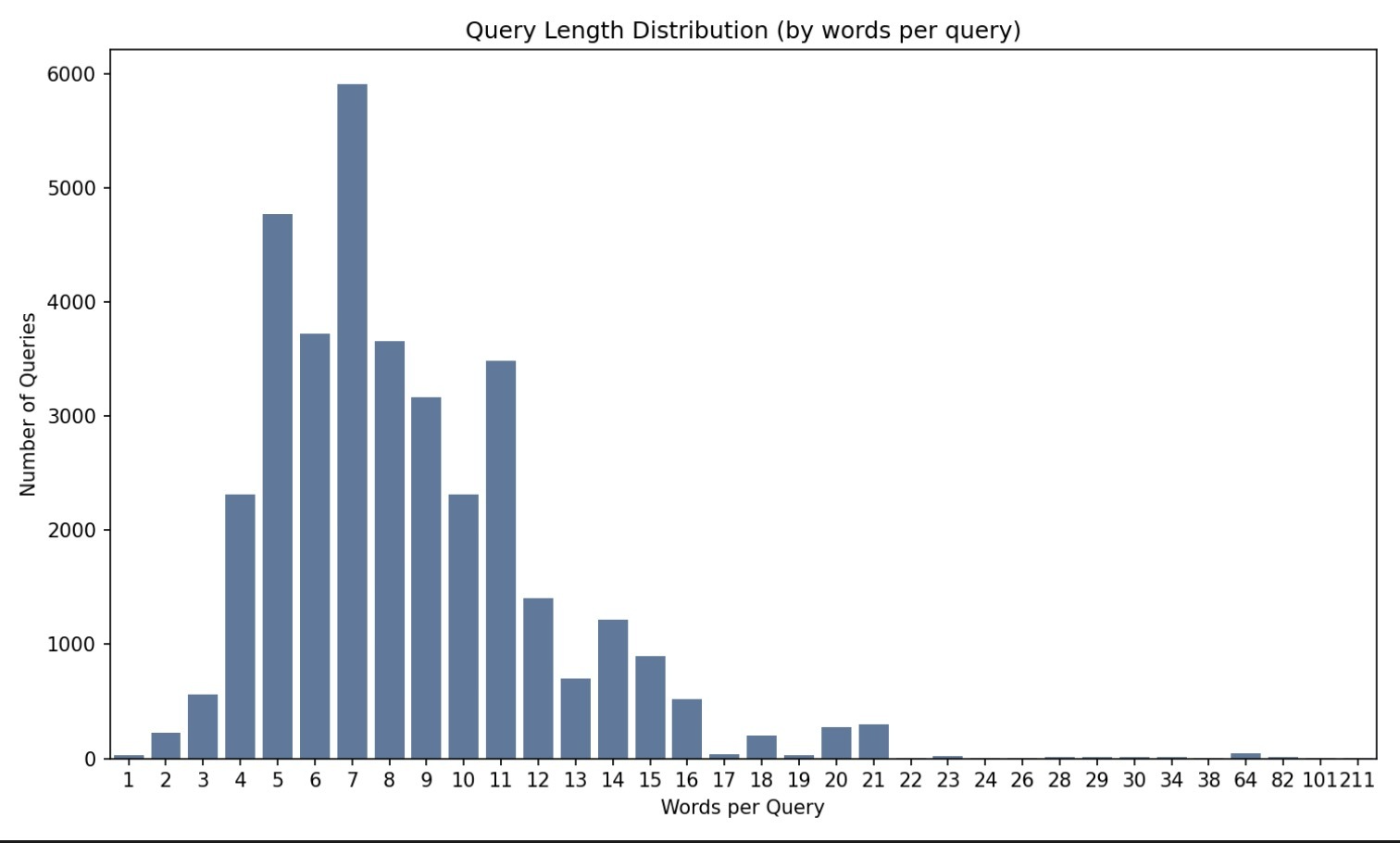}
\hfill
\includegraphics[width=0.48\textwidth]{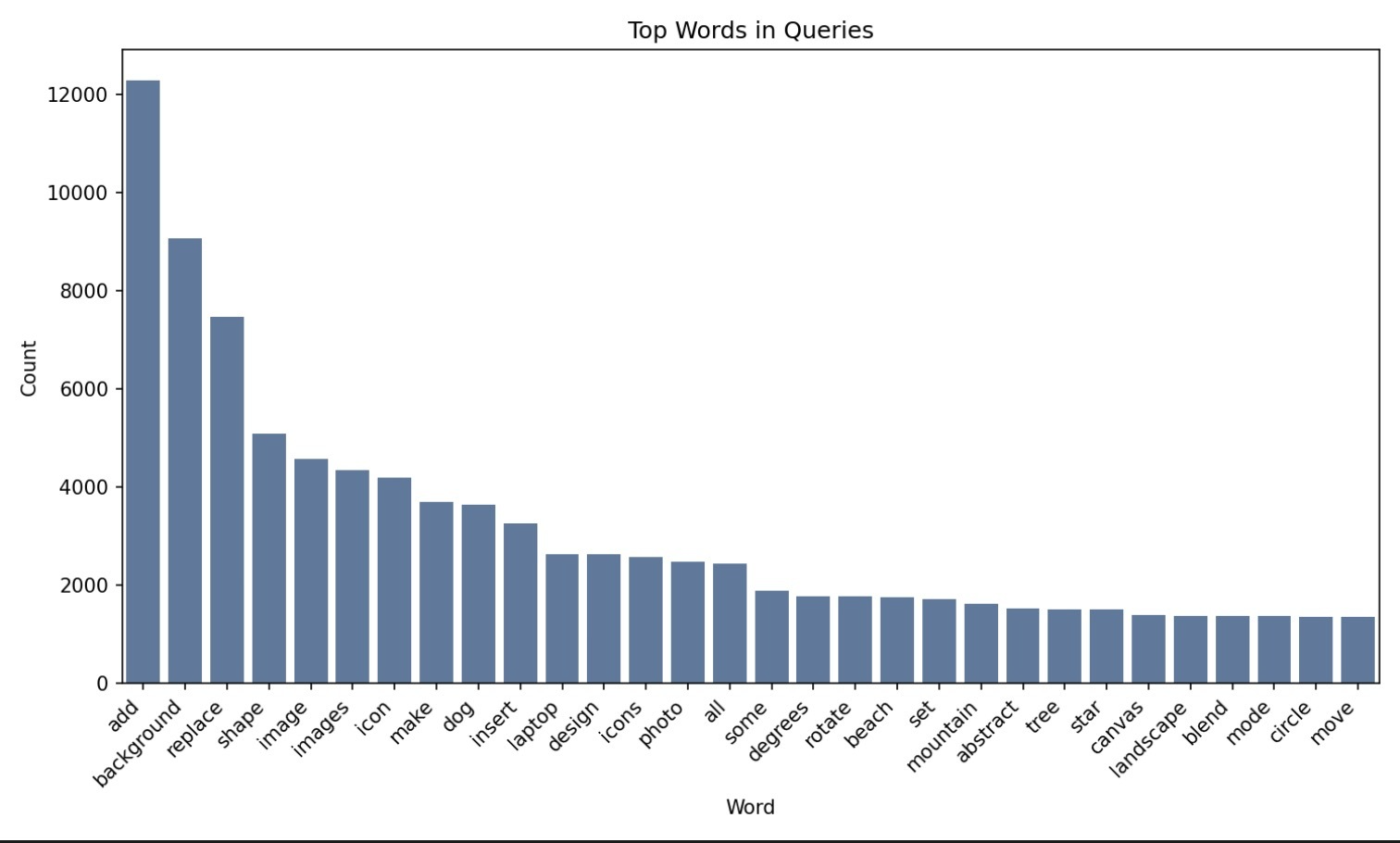}
\caption{Dataset created based out of large scale human evaluation. The left image shows query length distribution with the average length of seven words per query(after stemming). The right image shows the most frequently occurring words(after stemming) in the query, those being \emph{add}, \emph{background}, \emph{replace}, \emph{shape} and \emph{image}.}
\label{fig:prod_data_analysis}
\end{figure*}

\section{Related Work}

Within the image-rich project creation like brochures (refer Figure~\ref{fig:contextual_search}), search and recommendation extend beyond generic multimodal matching to incorporate template structure, visual style, and layout cues. Design-aware retrieval systems highlight the role of structured conditioning for template retrieval, while user-interactive creative assistants increasingly incorporate context signals such as selection history and semantic grouping. Despite these advances, existing systems primarily emphasize improving embedding quality or context fusion rather than systematically diagnosing where failures occur within multi-stage pipelines. Our approach addresses this gap by combining a compact Graphical Design Representation (GDR) with stage-wise failure attribution. This enables both efficient multimodal representation and systematic error diagnosis in production grade creative workflows.

\subsection{Agent-Orchestrated Search Systems}

Agent-orchestrated architectures have evolved from monolithic systems toward modular, specialized components coordinated by a Planner. Early frameworks such as ReAct~\cite{yao2023react} and Toolformer~\cite{schick2023toolformer} showed how large language models (LLMs) can interleave reasoning with external actions or teach themselves to invoke tools. Cascades~\cite{liu2023cascades} extended orchestration by recursively summarizing long contexts, while surveys of augmented language models highlight general strategies for tool use and memory integration. More recent production grade frameworks such as EHR-Agent~\cite{shi2024agent}, D-PoT for dynamic GUI planning~\cite{zhang2024dynamic}, and HM-RAG~\cite{liu2024hmrag} demonstrate effective coordination among specialized agents for query analysis, retrieval, and synthesis. However, these systems emphasize agent coordination efficiency or component-level optimization rather than systematic pipeline diagnosis. In contrast, our framework introduces a production grade attribution layer designed for low-latency, cost-aware multimodal environments. By operating under strict efficiency budgets, it provides systematic stage-wise diagnosis of failures across hierarchical agent interactions, enabling targeted optimization that is both actionable in practice and scalable beyond heuristic component tuning.

\subsection{Context Budgeting and Multimodal Understanding}

Context budgeting extends beyond simple prompt optimization to encompass broader information management in AI systems~\cite{schmid2025context,lutke2025context}. This field involves dynamic context construction, memory management, and tool integration. Vision-language models and multimodal frameworks~\cite{chen2024mllm} show growing capability in handling complex contexts. While existing context budgeting work focuses on general information architecture and context synthesis, our approach specifically targets context management in production grade multi-agent search systems, emphasizing design template constraints and reliable fallback handling.

\subsection{Pipeline Attribution and Error Analysis}

Traditional pipeline optimization relies on individual component tuning without systematic failure diagnosis~\cite{mazumder2020failure, ucar2024comprehensive}. Recent evaluation frameworks employ LLM-as-a-Judge methods and established metrics like nDCG and MRR~\cite{thomas2024improving,wang2024survey} for assessing search quality. However, existing attribution techniques focus on model-level explanations rather than stage-specific failure detection in complex multi-agent systems. Our framework advances beyond current approaches by introducing quantitative thresholds for systematic attribution across five pipeline stages, enabling precise identification of failure sources within agent-orchestrated architectures where traditional methods fail to distinguish between coordination failures and component-specific issues.

\section{System Architecture and Pipeline Attribution Framework}

We analyzed approximately 36,000 production-grade queries (shown in Figure \ref{fig:prod_data_analysis}) and found that most are short commands for asset operations—adding, replacing, or finding images, icons, or text styles. Crucially, many requests are context-dependent: creators frequently issue location-based or selection-dependent queries (e.g., ``recommend a background'' while a region is selected; ``replace this icon'' referring to a highlighted layer; ``find beans'' on a coffee poster). When executed as raw text, these queries omit crucial context—the target, canvas constraints (layout, aspect ratio, palette/typography, brand requirements)—leading to systematic failures: wrong content types, semantic mismatches (e.g., ``beans'' returning legumes instead of coffee beans), and layout incompatibilities. We manually evaluated a 788-case sample covering add/replace/find tasks from this evaluation dataset. Each case was judged for action correctness, asset relevance and layout/style coherence. Empirically, the raw query mode (no additional context) performs worst (63\% pass rate). These findings strengthen our experiment choice: infer the user’s intent (add/replace/find), resolve the target scope from the selection and object graph, and rewrite the raw query into a contextualized request that injects canvas semantics and feasibility constraints before retrieval and ranking. In short, relying on raw queries is disastrous in real creative workflows because users naturally ask context‑laden questions; robust systems must make that context explicit before they search.

\begin{figure*}[t]
  \centering
  \includegraphics[width=0.8\linewidth]{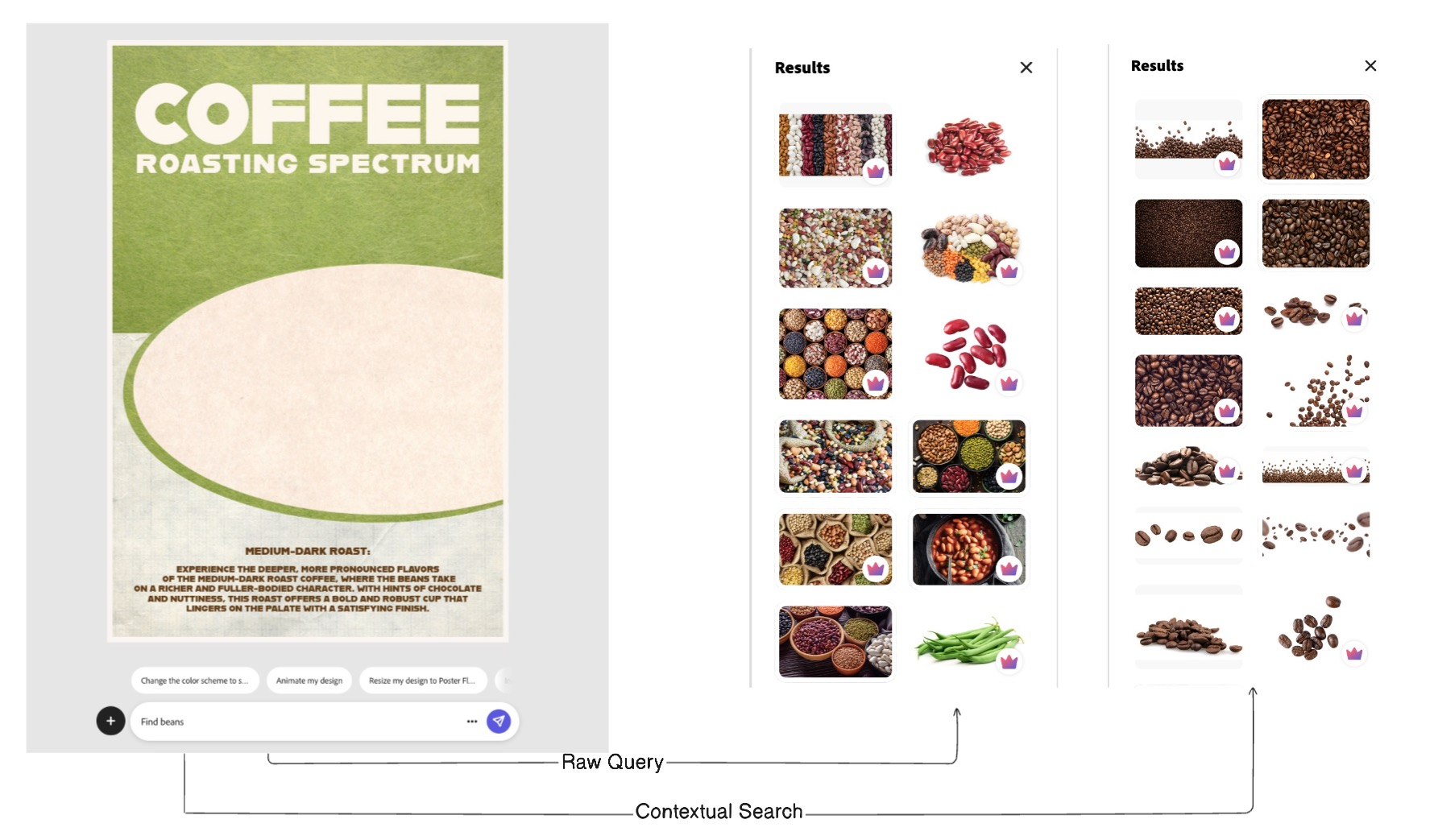}
  \caption{Example of context-aware search in a creative design workflow.  
A user working on a coffee roasting poster issues the query \emph{``find beans"}.  
The raw query returns generic beans (left result panel), while the contextual search:   
informed by the template content and design context prioritizes coffee beans (right result panel),  
demonstrating how contextual grounding improves retrieval relevance.}
  \label{fig:contextual_search}
\end{figure*}

The Planner model (responsible for routing to different subagents) infers context directly from the raw user query, using canvas as the hint, and then forward it as a free-form superagent prompt. If we would have used it directly
for recall, this could have been brittle: resulting in long,
redundant rewrites. Beyond quality concerns, the approach would have been operationally costly. Re-indexing and recalculating the embeddings of the massive corpus, comprising approximately 80 billion assets, would have been necessary. We instead moved to a more compact, typed contextual
schema (intent + canvas constraints) that drives retrieval/
ranking with explicit contrast
and feasibility checks. An example of context-aware search is shown in figure~\ref{fig:contextual_search}.

\subsection{Problem Formulation}
\noindent \textbf{Canvas:} a design state \(C\) with objects \(O=\{o_k\}\) (text, images, shapes), styles \(S\) (palette, typography), layout metadata (bounding boxes, z-order), and domain cues (e.g., ``coffee roasting spectrum'').

\noindent \textbf{User query:} an imperative text \(q\) (e.g., ``find beans") 

\noindent \textbf{Intent:} \(i \in \{\text{add},\text{replace},\text{find},\text{edit},\text{style}\}\) with target scope \(t \in \{\text{selection},\text{region},\text{global}\}\).

\noindent \textbf{Rewriting:} \(f(q,C,i,t)\mapsto q'\) injects semantic and geometric constraints (topic, aspect ratio, composition, licensing).

\noindent \textbf{Retrieval:} \(R(q')\) over dual indices (semantic vectors and lexical/metadata), returning \(K\) candidates.

\noindent \textbf{Re-ranking:} \(g\!\left(C,i,t,\{a_j\}_{j=1}^{K}\right)\mapsto \text{ranked list}\) via layout fit, brand/style similarity, and action feasibility.

\noindent \textbf{Execution:} the action planner selects top candidate(s) and applies a transformation consistent with \(i,t\).

\subsection{Canvas grounding and Intent inference}

 As shown in Figure~\ref{fig:architecture}, we use a lightweight GPT‑4o‑mini model for intent understanding and the canvas has llava descriptions of the asset. The canvas is serialized as a \emph{Graphical Design Representation (GDR)}: a typed JSON that records each page and element with geometry and appearance (opacity, zIndex), placement (tx, ty, rotation, scales), precise frame coordinates (topLeft, topRight, bottomRight, bottomLeft), color roles and palettes, and LLaVA captions plus detected regions. Example of GDR is presented below: 
 
\begin{lstlisting}[frame=none, caption=Example of Graphical Design Representation]
{
  "gdrVersion": "assistant-100",
  "pages": [{
    "id": "page-0",
    "elements": [{
      "id": "image-0", "type": "image", "opacity": 1, "zIndex": 1,
      "frame": { "topLeft": [x0,y0], "bottomRight": [x1,y1] },
      "colors": [{"color":"#040404","weight":0.71}, ...],
      "content": ["A blue background with vegetables ..."],
      "regions": [{ "label":"octopus (animal)", "score":0.79, ... }]}]}]
}
\end{lstlisting}
\label{lst:gdr_representation}

At query time, GPT-4o-mini takes the GDR and planner prompt to create a detailed subprompt that makes hidden design requirements explicit. This process extracts six key elements: what action to take (add/replace/find), where to apply it (selected item, region, or whole canvas), size and shape needs (aspect ratio, dimensions), color compatibility (matching palettes, ensuring contrast), style consistency (textures, brand alignment), and content rules (licensing, appropriateness).

This approach fixes common problems with basic contextual search. For instance, if someone searches for ``find similar cars" while working on a design with a green background, a regular contextual search might return green cars that disappear into the background. FUSE's subprompt instead specifies ``find cars with sufficient contrast against green" and checks that results will actually be visible when placed on the canvas.

Our focus centers on the search subagent ecosystem that implements a comprehensive five-stage pipeline: pre-processing, intent understanding, routing, recall, and ranking. Unlike distributed architectures employing multiple search subagents, our system utilizes a unified search subagent capable of processing queries across multiple content surfaces including stock photos, templates, enterprise assets, and multimedia content. The search subagent connects to specialized Entity Search Handlers that interface with domain-specific ElasticSearch indices (e.g. for photos, icons, backgrounds, video clips, audio) enabling comprehensive multimodal searches while maintaining optimized retrieval performance through distributed indexing

While Planner-provided GDR‑grounding with a GPT‑4o‑mini intent model eliminated most raw‑query failures, end‑to‑end latency on editors still hovered at 2.8\,s (p95), driven largely by an oversized system prompt and heavy few‑shot conditioning. Much of the injected context did not apply uniformly across asset classes; conflicting hints in the prompt confused retrieval and ranking (e.g., adding shape‑specific cues when our index had sparse shape coverage frequently yielded zero hits). To resolve this, we redesigned context handling to express canvas constraints explicitly \emph{without} long prompts and to move expensive steps off the critical path.

\subsection{Context Budgeting}

We conducted comprehensive experiments across seven context budgeting variants, each testing different information density and processing strategies:
\begin{itemize}
\item{\textbf{Baseline Configuration}}: employs comprehensive system prompts with full few-shot examples across all content types. While maximizing available information, this ``context-rich" approach often overwhelms the model with irrelevant details and incurs substantial token costs, particularly when processing complex canvas states.

\item{\textbf{Chain-of-Thought Contextualization}} augments standard context with intermediate reasoning traces through an assistant ``scratchpad." The model explicitly reasons through intent inference and constraint extraction before generating subprompts, improving complex query handling at increased token cost.

\item{\textbf{Context Compression}} sends the summarized version to compress GDR representations and prompts while maintaining semantic richness, testing whether information density improvements translate to better performance-cost ratios. 

\item{\textbf{Mini-Shot Optimization}} provides one strategically curated example per content type, testing whether targeted context selection outperforms volume-based approaches.

\item{\textbf{Retrieval-Augmented Context}} dynamically selects few-shot examples by computing embedding similarity between current queries and historical examples. This ensures contextual relevance while maintaining computational efficiency through asset-specific example banks.

\item{\textbf{Two-Stage Hierarchical Processing}} first predicts content type, then applies asset-specific context conditioning. Once the system predicts ``shapes", it skips contextual examples entirely given limited inventory coverage, while photo queries receive rich conditioning. This reduces context noise while preserving semantic richness where beneficial.

\item{\textbf{Zero-Shot Minimalism}} strips context to bare system prompts without guidance examples, establishing performance floors while revealing inherent model capabilities. This ``context-lean" baseline processes only essential GDR elements and planner prompt.
\end{itemize}
Each variant is systematically evaluated through our Pipeline Attribution Framework, which tracks failure distribution across the five stage pipeline: pre-processing, intent understanding, routing, recall, ranking. 

\subsection{Pipeline Attribution Framework}

To evaluate pipeline performance, we developed Pipeline Attribution Framework that classifies failures into four distinct attribution categories a) Intent b) Routing c) Recall and d) Ranking by comparing predicted outputs against expected ground truth at each pipeline stage respectively.

Intent Attribution captures failures in extracting user intent from natural language queries. Example: For the query ``add photo of beans" we expect subprompt to be ``coffee beans" due to the GDR context (template has coffee elements in it) but the system predicted ``black beans" as the main search subprompt, indicating complete semantic misunderstanding between user intent and system interpretation.

Routing Attribution identifies incorrect agent selection decisions despite successful intent extraction. Example: A system correctly identifies ``fireworks" intent but incorrectly routes it to video content type search instead of the expected photo content type search, requiring fallback routing mechanisms. Having incorrect routing can potentially result in change in the asset modality and degrade user experience.

Recall Attribution indicates content coverage gaps where the system fails to retrieve any relevant results due to a lack of assets. Example: The query ``create a deer pattern background" returns zero hits, suggesting missing content in the underlying database or indexing failures.

Ranking Attribution captures cases where relevant content is retrieved but incorrectly ordered by relevance. Example: For the query ``stone pattern backgrounds" if the most semantically similar result (similarity 0.674) appears at rank 3 and a less relevant result (similarity 0.474) occupied rank 1 in the search results list.

\begin{figure*}[t]
    \centering
    \includegraphics[width=0.48\textwidth]{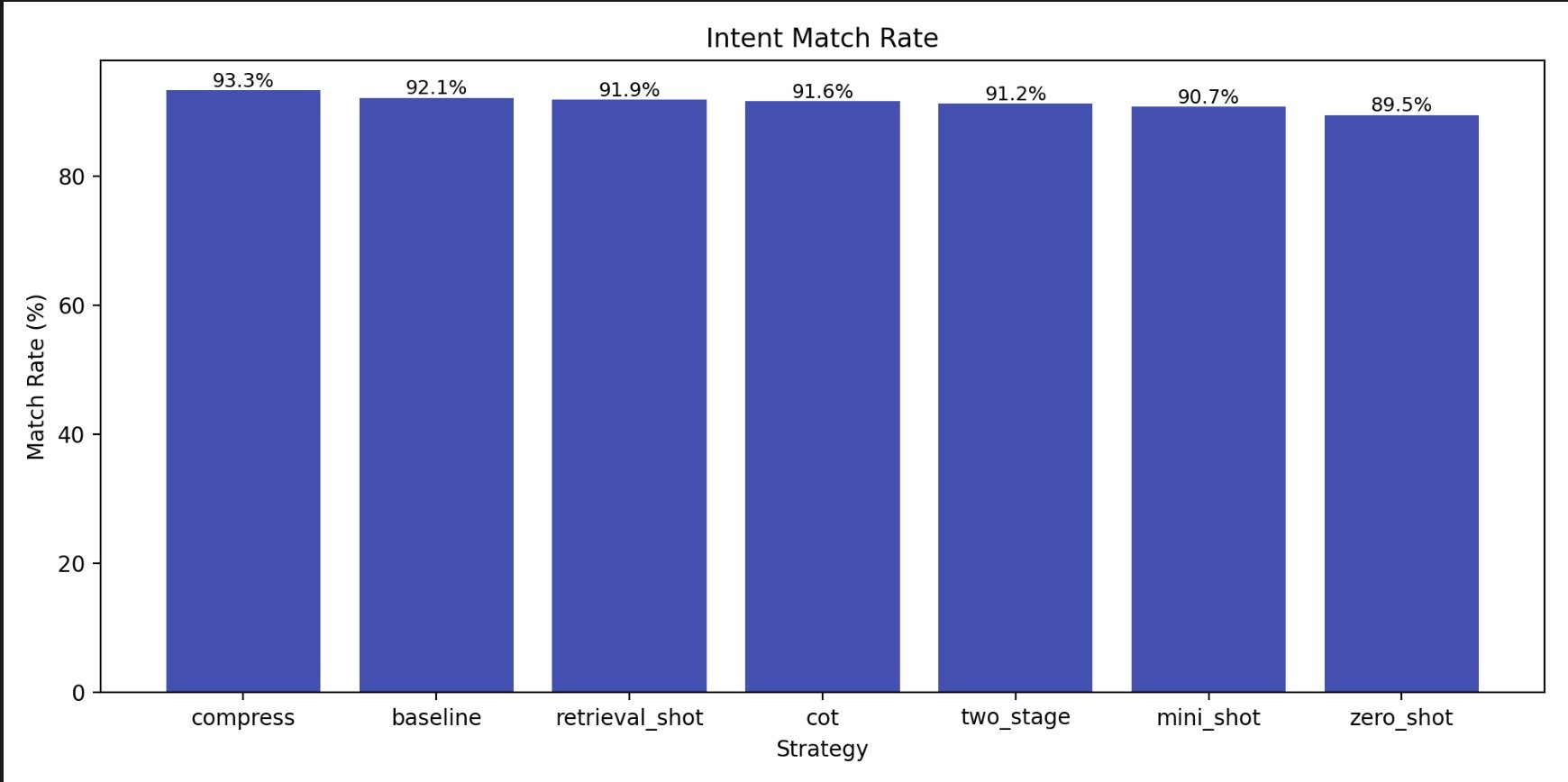}
    \includegraphics[width=0.48\textwidth]{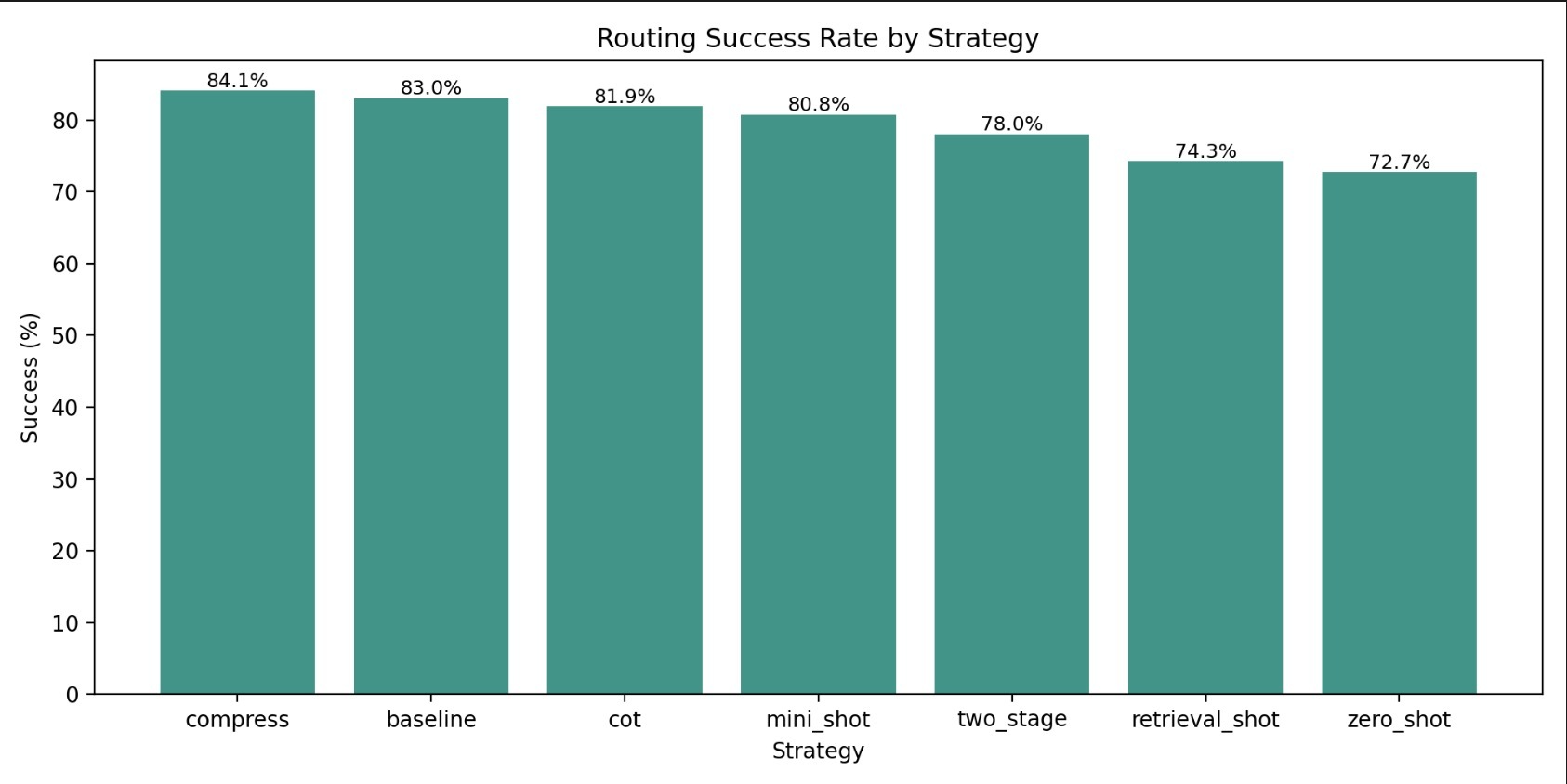}
    \includegraphics[width=0.48\textwidth]{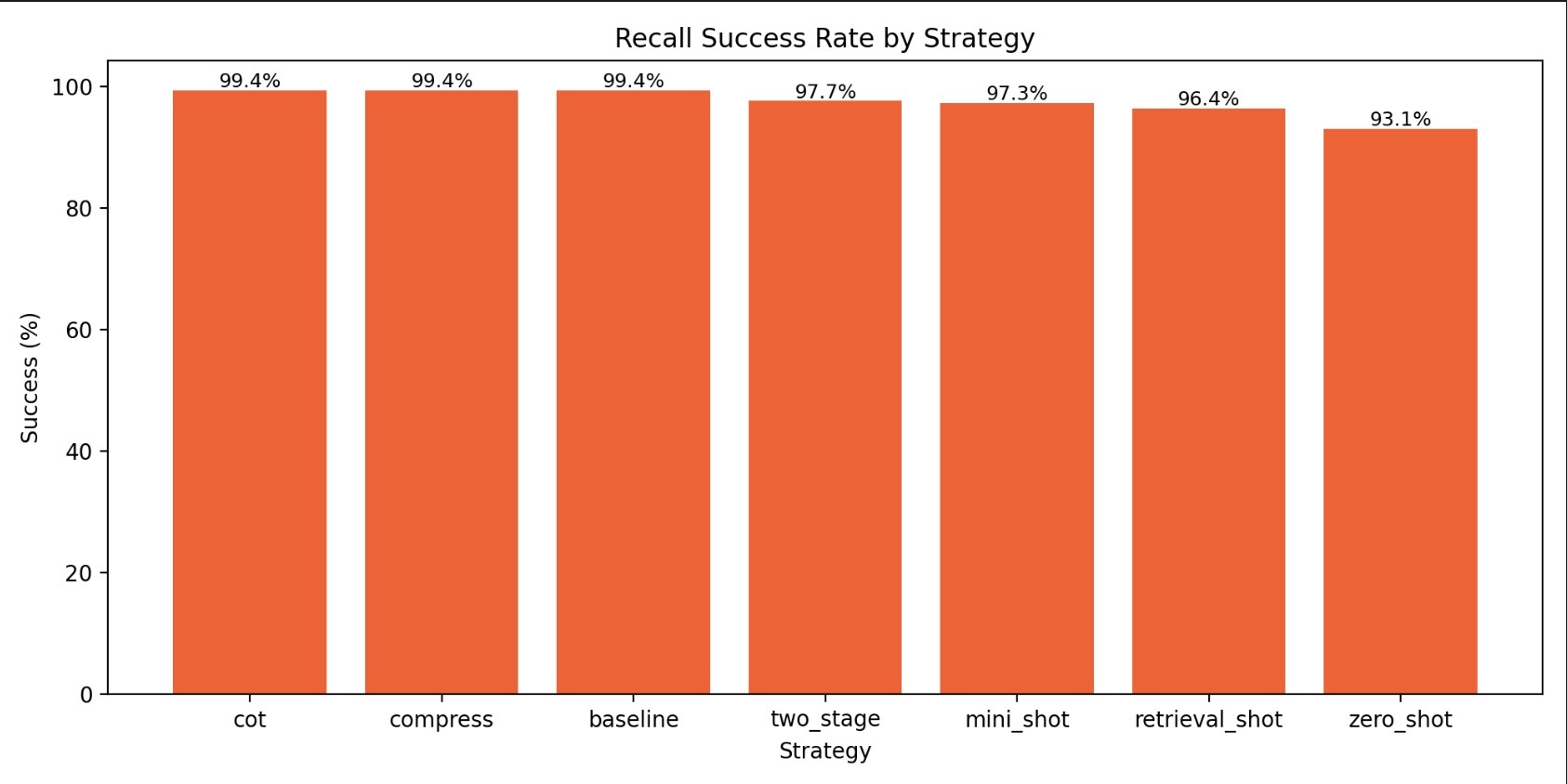}
    \includegraphics[width=0.48\textwidth]{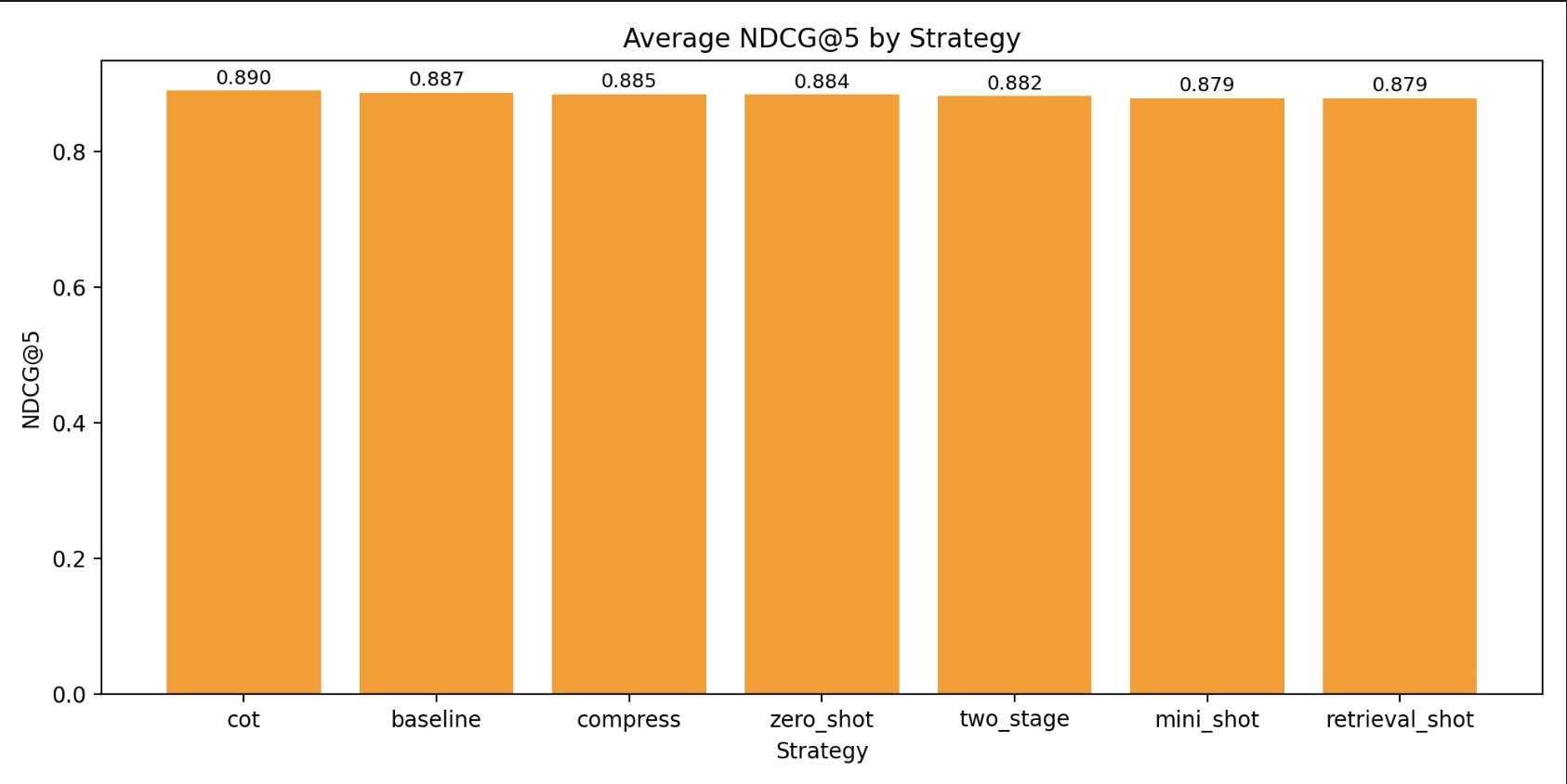}
    
    \caption{We analyze Pipeline Performance for \emph{Intent Match Rate}, \emph{Routing Success Rate}, 
    \emph{Recall Success Rate} and
    \emph{Average NDCG@5} across Context Budgeting strategies}
    \label{fig:results}
\end{figure*}

\subsection{Evaluation Metrics}

We employ stage-specific metrics aligned with our Performance Attribution Framework to evaluate pipeline performance across the seven context budgeting variants.

Intent Evaluation counts semantic mismatches (below a certain threshold) between ground truth subprompts and system-predicted subprompts. For example, the query ``replace image with a modern salon" expects the subprompt ``modern salon" - successful intent extraction increments the match count, while mismatched predictions like ``dandelions" contribute to failure counts.

Routing Evaluation counts exact matches between expected content type and predicted content type. Queries expecting ``photo" content type that correctly predict ``photo" increment success counts, while routing failures like predicting ``icon" instead of ``shape" increment failure counts.

Recall Evaluation counts zero-hit queries that return no search results. Queries like ``create a deer pattern background" returning empty result sets increment the zero-hit count, indicating recall failures and content coverage gaps.

Ranking Evaluation employs NDCG@5 to assess result ordering quality within the top 5 positions, reflecting user focus on initial search results. NDCG@5 uses semantic similarity scores as relevance judgments, rewarding systems that place higher-similarity results at top ranks. For instance, a query where the best semantic match appears at rank 5 instead of rank 1 receives lower NDCG@5 scores than optimal ordering.

\begin{figure*}[t]
    \centering
    \includegraphics[width=\textwidth]{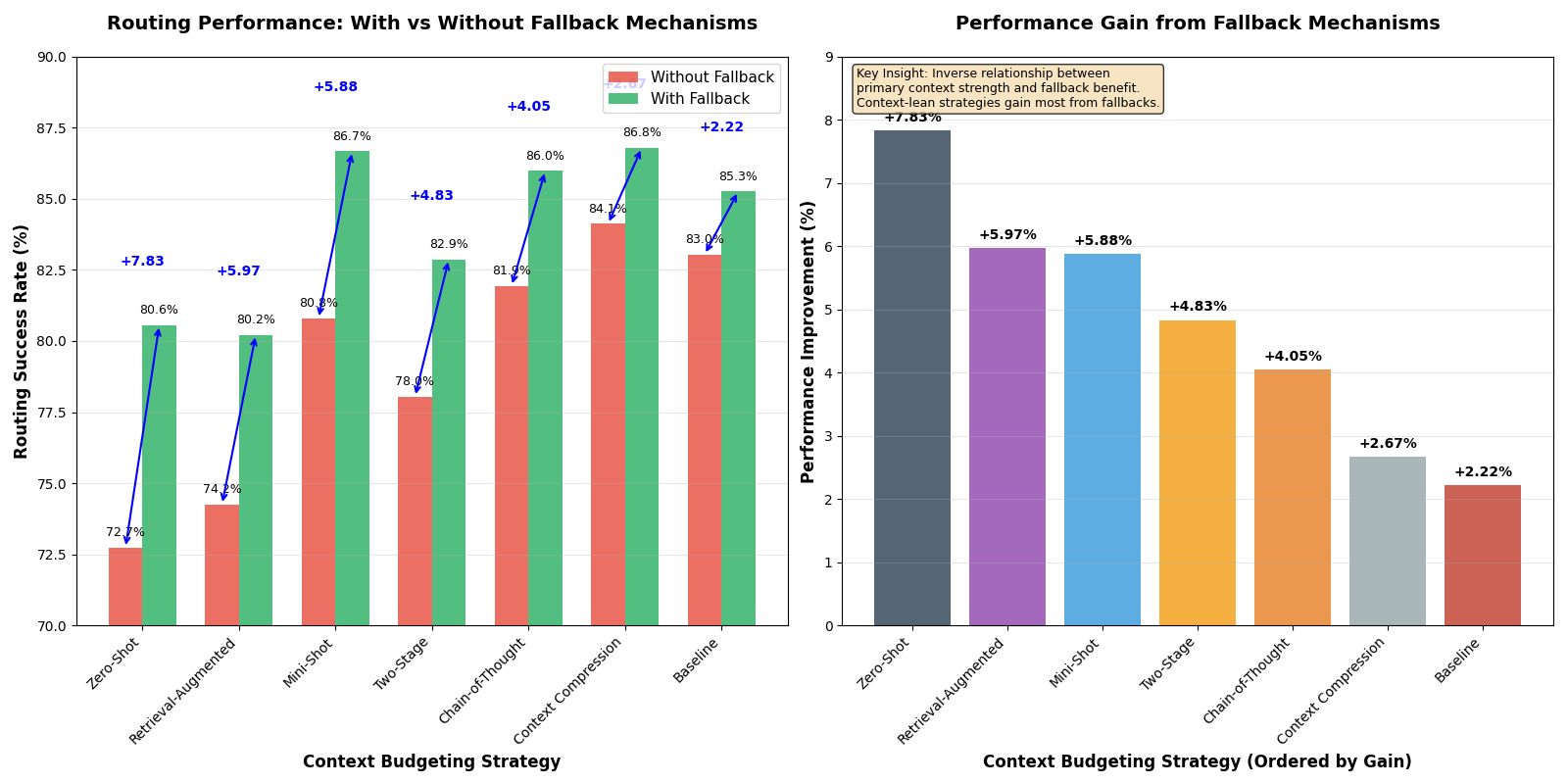}
    \caption{Improvement in routing success rate with fallback across Context Budgeting strategies. The strategies are ordered by most gain in performance.}
    \label{fig:results_routing_with_fallback}
\end{figure*}

\section{Results and Analysis}

\begin{table}[h]
\centering
\caption{Comprehensive Performance Analysis Across Context Budgeting Strategies}
\label{tab:results_tabular}
\scriptsize
\begin{tabular}{l|ccccc}
\toprule
{\textbf{Variant}} & \multicolumn{5}{c|}{\textbf{Performance Metrics (\%)}} \\
 & \textbf{Intent} & \textbf{Routing} & \textbf{Recall} & \textbf{NDCG@5} & \textbf{Overall} \\
  & \textbf{Match} & \textbf{Success} & \textbf{Success} & \textbf{Success} \\
\midrule
Baseline & 92.1 & 83 &  \textbf{99.4} & 88.7 & 80.8 \\
Chain-of-Thought & 91.6 & 81.9 &  \textbf{99.4} & \textbf{89.0} & 81.7 \\
Context Compression & \textbf{93.3} & \textbf{84.1} & \textbf{99.4} & 88.5 & \textbf{82.9} \\
Mini-Shot  & 90.7 & 80.8 & 97.3 & 87.9 & 78.6 \\
Retrieval-Augmented & 91.9 & 74.3 & 96.4 & 87.9 & 79.4 \\
Two-Stage  & 91.2 & 78 & 97.7 & 88.2 & 79.9 \\
Zero-Shot  & 89.5 & 72.7 & 93.1 & 88.4 & 74.5 \\
\bottomrule
\end{tabular}
\end{table}

\begin{figure*}[t]
\centering
\begin{tikzpicture}
\begin{axis}[
  ybar=0pt,
  bar width=10pt,
  width=\textwidth,
  height=7cm,
  ymin=0,
  ylabel={Latency (ms)},
  symbolic x coords={Baseline,Chain-of-Thought,Context Compression,Mini-Shot,Retrieval-Augmented,Two-Stage,Zero-Shot},
  xtick=data,
  xticklabel style={rotate=35, anchor=east},
  legend style={at={(0.5,1.15)},anchor=south,legend columns=-1},
  ymajorgrids=true,
  nodes near coords,
  nodes near coords align={vertical}
]

   \addplot coordinates {
    (Baseline,1560)
    (Chain-of-Thought,1790)
    (Context Compression,860)
    (Mini-Shot,1400)
    (Retrieval-Augmented,1250)
    (Two-Stage,1530)
    (Zero-Shot,1010)
  };
  \addplot coordinates {
    (Baseline,2793)
    (Chain-of-Thought,3210)
    (Context Compression,1540)
    (Mini-Shot,2510)
    (Retrieval-Augmented,2230)
    (Two-Stage,2740)
    (Zero-Shot,1820)
  };
  \addplot coordinates {
    (Baseline,4145)
    (Chain-of-Thought,4770)
    (Context Compression,2280)
    (Mini-Shot,3730)
    (Retrieval-Augmented,3320)
    (Two-Stage,4060)
    (Zero-Shot,2690)
  };
  \legend{p50, p95, p99}

  \end{axis}
  \begin{axis}[
    width=\textwidth,
    height=7cm,
    axis y line*=right,
    axis x line=none,
    ymin=0, ymax=1.1,
    ylabel={Relative cost (tokens/query)},
    symbolic x coords={Baseline,Chain-of-Thought,Context Compression,Mini-Shot,Retrieval-Augmented,Two-Stage,Zero-Shot},
    xtick=data,
  ]
  
    \addplot+[mark=*, color=red] coordinates {
      (Baseline,0.90)
      (Chain-of-Thought,1.00)
      (Context Compression,0.12)
      (Mini-Shot,0.70)
      (Retrieval-Augmented,0.50)
      (Two-Stage,0.88)
      (Zero-Shot,0.35)
    };
    
  \end{axis}
\end{tikzpicture}
\caption{Latency distribution (p50/p95/p99) with relative cost overlay across contextualization strategies.}
\label{fig:latency_cost}
\end{figure*}
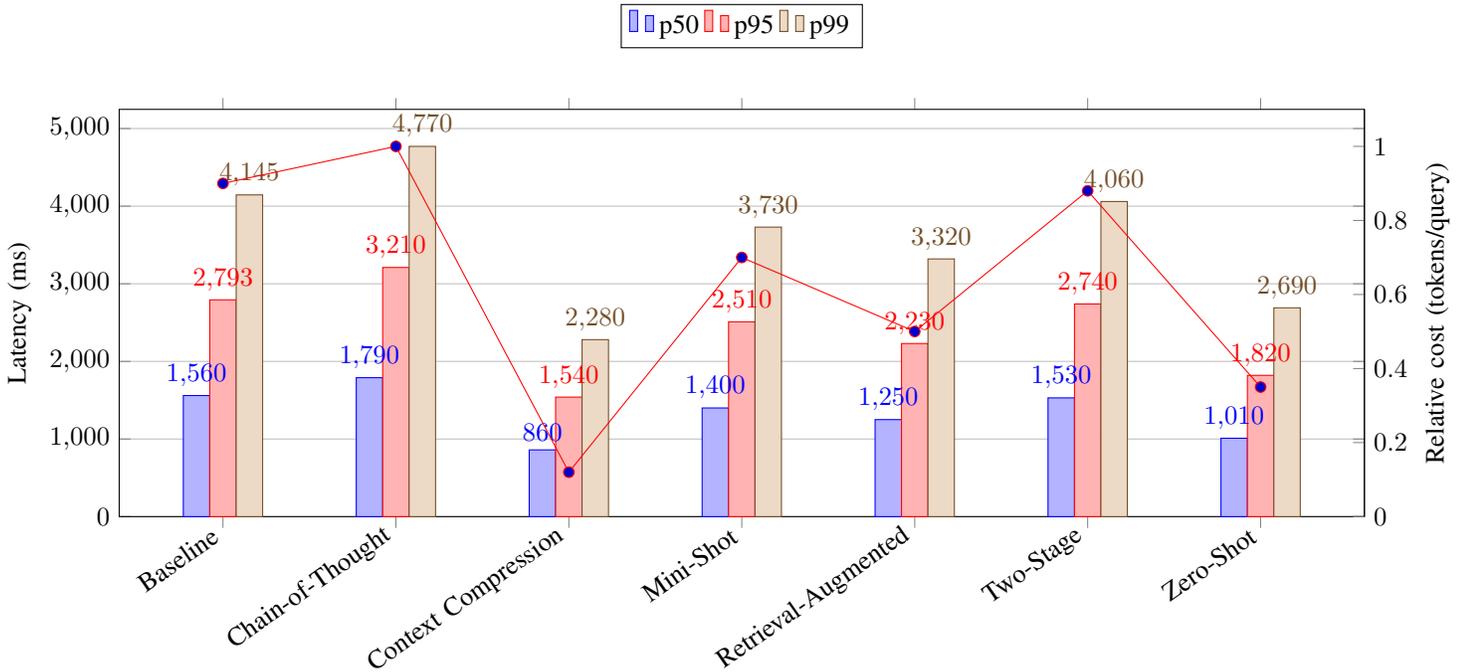

Our systematic evaluation of 788 evaluation (templates + queries) across seven context budgeting variants reveals distinct performance patterns at each pipeline stage. Of these we only consider scenarios where there is no failure on the planner side and the requests reach the search subagent. The error attribution analysis demonstrates that different context strategies produce markedly different failure distributions, validating our hypothesis that context engineering impacts vary significantly across pipeline components.

\subsection{Intent Extraction Performance}
Intent extraction demonstrates consistent performance across context budgeting variants, ranging from 89.5\% to 93.3\% accuracy. Context Compression achieves optimal performance at 93.3\%, followed by Baseline (92.1\%) and Retrieval-Augmented (91.9\%). Chain-of-Thought (91.6\%), Two-Stage (91.2\%), and Mini-Shot (90.7\%) maintain competitive accuracy, while Zero-Shot (89.5\%) establishes a minimal context baseline(Table~\ref{tab:results_tabular})

The narrow 3.8-percentage-point variance indicates that all context budgeting strategies provide substantial semantic understanding benefits, with even minimal approaches achieving production-viable accuracy. This validates our hypothesis that strategic context engineering enhances intent extraction while creating opportunities for computational optimization.

\subsection{Routing Performance and Fallback Mechanisms}

Routing performance exhibits the highest variance among pipeline stages, ranging from 72.7\% to 84.12\% success rates. Context Compression achieves optimal routing performance at 84.12\%, followed by Baseline (83.04\%).

Retrieval-based strategies demonstrate routing challenges, with Zero-Shot (72.73\%) and Retrieval-Augmented (74.25\%) exhibiting the lowest success rates. This 12-percentage-point variance confirms that content type prediction requires stable contextual guidance, with curated compression approaches outperforming dynamic retrieval methods for routing decisions.

\subsection{Fallback Mechanism Enhancement}

\begin{table}[h]
\centering
\caption{Performance Improvement Using Fallbacks Across Context Budgeting Across}
\label{tab:results_tabular_fallback}
\scriptsize
\begin{tabular}{l|ccc}
\toprule
{\textbf{Variant}} & \multicolumn{3}{c}{\textbf{Performance Metrics (\%)}} \\
 & \textbf{Without} & \textbf{With} & \textbf{Performance} \\
  & \textbf{Fallback} & \textbf{Fallback} & \textbf{Gain} \\
\midrule
Baseline & 83.04 & 85.26 &+2.22 \\
Chain-of-Thought & 81.93 & 85.98 & +4.05 \\
Context Compression & 84.12	& 86.79 & +2.67 \\
Mini-Shot  & 80.79 & 86.67 & +5.88 \\
Retrieval-Augmented & 74.25 & 80.22 & \textbf{+5.97} \\
Two-Stage  & 78.04 & 82.87 & +4.83 \\
Zero-Shot  & 72.73 & 80.56 & +7.83 \\
\bottomrule
\end{tabular}
\end{table}

Fallback routing mechanisms provide consistent performance improvements across all variants as demonstrated in Table ~\ref{tab:results_tabular_fallback} and Figure~\ref{fig:results_routing_with_fallback}, with gains ranging from 2.22 to 7.83 percentage points. Context-lean strategies benefit most dramatically: Zero-Shot (+7.83 points), Retrieval-Augmented (+5.97 points), and Mini-Shot (+5.88 points) show that robust fallback logic compensates for limited primary contextualization.

Context-rich strategies demonstrate modest but consistent gains, with Compression (+2.67 points) and Baseline (+2.22 points) showing reduced fallback dependency. This inverse relationship indicates that strategies with stronger primary context achieve consistency through pathway optimization, while context-lean approaches rely more heavily on fallback recovery mechanisms.

\subsection{Ranking Quality Assessment}

Ranking performance measured via NDCG@5 demonstrates notable consistency across context budgeting variants, with scores ranging from 87.9\% to 89\% (Table~\ref{tab:results_tabular}). This 0.011\% variance represents stable performance across different contextualization approaches.

Chain-of-Thought Context leads ranking performance at 0.890 NDCG@5, followed closely by Baseline (88.7\%) and Context Compression (88.5\%). The competitive performance across different approaches suggests that ranking algorithms benefit from but are not critically dependent on upstream contextualization strategies.

The ranking consistency indicates that semantic embeddings capture sufficient information for effective result ordering regardless of upstream context processing. This finding suggests that ranking optimization efforts should focus on embedding quality and similarity calculation methods rather than extensive context engineering.

\subsection{Recall Performance Analysis}

Recall performance demonstrates varied effectiveness across context budgeting variants, ranging from 93.1\% to 99.4\% success rates. Context-rich strategies achieve optimal performance, with Compression, Baseline, and Chain-of-Thought all reaching 99.4\% recall success, demonstrating comprehensive content coverage under well-contextualized conditions.

Moderate context strategies show manageable degradation: Two-Stage (97.7\%) and Mini-Shot (97.3\%) maintain acceptable production grade thresholds, while Retrieval-Augmented (96.4\%) and Zero-Shot (93.1\%) exhibit more significant challenges. The 6.3-percentage-point variance indicates that minimal context strategies require enhanced retrieval algorithms to maintain comprehensive coverage.

\subsection{Overall Success Analysis}

Context Compression emerges as the superior strategy across all pipeline stages, achieving 93.3\% intent accuracy, 86.8\% routing success, 99.4\% recall, and 88.5\% NDCG@5. This approach leverages planner context summarization to generate compact, semantically-rich representations that preserve critical information while reducing prompt complexity.

Figure~\ref{fig:latency_cost} shows p50/p95/p99 end-to-end latency across strategies. Context Compression is fastest (p95 1.54\,s), followed by Zero-Shot (1.82\,s), Retrieval-Augmented (2.23\,s), Mini-Shot (2.51\,s), Two-Stage (2.74\,s), Baseline (2.79\,s), and Chain-of-Thought (3.21\,s). The ordering aligns with context size: shorter prompts (compression/zero-shot) reduce serialization and model time; adding few-shots or reasoning increases tail latency. Tails widen most under Chain-of-Thought, reflecting added reasoning tokens.

We compute relative cost per strategy as proportional to total tokens processed by the GPT-4o-mini intent model, normalized to Chain-of-Thought = 1.00: \( \mathrm{rel\_cost}(s) = \tfrac{\mathrm{tokens\_in}(s) + \mathrm{tokens\_out}(s)}{\mathrm{tokens\_in}(\mathrm{CoT}) + \mathrm{tokens\_out}(\mathrm{CoT})} \). Token counts include: (i) the fixed system instructions in SystemPrompt, (ii) strategy-specific few-shots from GPTShots, and (iii) conext from planner (planner prompt + GDR JSON). Output tokens are small and included for completeness. For reference, a 7-word output (~9 tokens) costs ~5–6 micro-dollars at \$0.60 per 1M output tokens.

Given these results, we recommend Context Compression as the production grade strategy, as it consistently delivers optimal performance across all pipeline components while providing a scalable framework for context management in large-scale deployment scenarios.

\section{Conclusion}

This work presents the first comprehensive study of context budgeting strategies in production-grade multimodal search and recommendation systems, addressing a critical gap between research-oriented models and deployment-oriented constraints. Through the systematic evaluation of seven distinct context budgeting variants across 788 real-world templates and evaluation queries, we demonstrate that context engineering decisions exert a decisive influence on end-to-end system performance, with their impact varying significantly across pipeline stages. Empirical results reveal that context-rich strategies, while incurring higher computational cost, yield substantial gains in query understanding and user experience, exceeding 50-point performance improvements over minimal-context baselines, thus justifying their adoption in production-grade, user-facing applications. These findings underscore that context budgeting cannot be treated as a local optimization problem; rather, it requires a holistic, pipeline-aware evaluation using production relevant metrics and real user data. The proposed Performance Attribution Framework, together with the experimental insights, offers actionable guidance for practitioners seeking to optimize multimodal retrieval systems under real-world constraints, enabling principled trade-offs between computational efficiency and user experience fidelity.

Operationally, the cost overlay in Figure~\ref{fig:latency_cost} (right axis) confirms that Context Compression minimizes token spend while preserving quality, delivering the strongest latency–cost trade-off; Chain-of-Thought provides the most reasoning headroom at the highest cost.




\bibliography{references}
\bibliographystyle{IEEEtran}

\end{document}